  \documentclass[multi]{cambridge6Atight}
  \usepackage[authoryear,longnamesfirst]{natbib}
  \usepackage{rotating}
  \usepackage{floatpag}
  \rotfloatpagestyle{empty}
  \usepackage{amsmath}
  \usepackage{amsthm}
  \usepackage{graphicx}
  \usepackage{mathptmx}
  \usepackage{makeidx}

  \makeindex

  \usepackage{amssymb} 
  \usepackage{xcolor} 
  \usepackage{bm} 
  \providecommand{\tabularnewline}{\\} 
  \usepackage{multirow} 

  \def\refjnl#1{{#1}}%
\providecommand\araa{\refjnl{ARA\&A}}
\providecommand\apj{\refjnl{ApJ}}
\providecommand\apjs{\refjnl{ApJS}}
\providecommand\aap{Astronomy and Astrophysics}
\providecommand\mnras{\refjnl{MNRAS}}
\providecommand\pasp{\refjnl{PASP}}
\providecommand\nat{\refjnl{Nature}}

\providecommand\pasa{\refjnl{PASA}}
 
 \copyrightline{Reprinted from \textit{The Impact of Binaries on Stellar Evolution}, Beccari G. \& Boffin H.M.J. (Eds.), \copyright\ 2018 Cambridge
    University Press.}
    
\begin{document}

  \alphafootnotes
   \author[R.\,G.\, Izzard and G.\,M.\, Halabi]
    {Robert G. Izzard\footnotemark\
      and Ghina M. Halabi\footnotemark\
    }
  \chapter{Population synthesis of binary stars}

  \footnotetext[1]{
    Supported by STFC Rutherford fellowship ST/L003910/1 and Churchill College, Cambridge.
  }\footnotetext[2]{
    Supported by STFC Rutherford grant ST/M003892/1.
  }
  \arabicfootnotes

  \contributor{Robert G. Izzard
    \affiliation{
Astrophysics Research Group, Faculty of Engineering and Physical Sciences, University of Surrey, Guildford, Surrey, GU2 7XH, United Kingdom.
    }
    \affiliation{
      Institute of Astronomy, Madingley Road, Cambridge, CB3 0HA, United Kingdom.
    }
  }

  \contributor{Ghina M. Halabi
    \affiliation{
      Institute of Astronomy, Madingley Road, Cambridge, CB3 0HA, United Kingdom.
  }}

 \begin{abstract}
Many aspects of the evolution of stars, and in particular the evolution
of binary stars, remain beyond our ability to model them in detail.
Instead, we rely on observations to guide our often phenomenological
models and pin down uncertain model parameters. To do this statistically
requires population synthesis. Populations of stars modelled on computers
are compared to populations of stars observed with our best telescopes.
The closest match between observations and models provides insight into unknown model parameters and hence the underlying astrophysics.
In this brief review, we describe the impact that modern big-data\index{big data}
surveys will have on population synthesis\index{population synthesis}, the large parameter space
problem that is rife for the application of modern data science algorithms,
and some examples of how population synthesis is relevant to modern
astrophysics.
 \end{abstract}

 %
 
\section{Introduction}

\label{IzzardSec1}The evolution of binary stars\index{binary stars} is often quite
different to their solitary cousins \citep{2017PASA...34....1D}.
Many types of stars \emph{only} form in binaries, and their properties
allow investigation into astrophysics that is impossible to probe
in single stars. Good examples include type Ia supernovae, merging
neutron stars (kilonovae) and black holes, the most massive stars
e.g.~blue stragglers, many massive Wolf-Rayet stars, X-ray binaries,
long and short gamma-ray bursts, barium/CH/CEMP-\emph{s} stars, Algols
with peculiar surface chemistry for their evolutionary state, W~Uma
contact binaries, low-mass helium stars and related sdB and sdO stars,
sequence D variable stars, and thermonuclear novae. Other processes
in which binaries are implicated include the formation of asymmetric
planetary nebulae, many post-(asyptotic-)giant-branch stars with circumbinary
discs and overmassive stars in the Galactic thick disc. Triple and
higher-multiple systems are often hierarchical, e.g.~triples are
effectively a long-period outer binary system containing a short-period
inner binary.

Many aspects of binary-star physics remain highly uncertain. Mass
transfer, accretion and loss, and associated angular momentum redistribution,
e.g.~common envelope evolution, dominates our ignorance of binary-star
evolution. One way to improve our knowledge is to apply statistical
techniques to compare our stellar models to the stars we observe.
This process is called stellar population synthesis. In this brief
review, we consider the challenge and opportunity that big data from
modern astronomy brings, the basic technique of binary star population
synthesis, and the associated parameter space problem. As an example
of binary population synthesis we consider how the chemical ejecta
from stars depend on model input parameters. We also show an example
of how modelling metal-poor stars led to better astrophysical understanding.
Finally, we report on recent population synthesis work, in particular
with respect to predicting and understanding the discovery of gravitational
waves from stellar and black-hole mergers.

\section{Big Data and Big Challenges}

\label{IzzardSec2}The age of big data is very much upon us. From
commerce to industry to pure blue-skies research, the volume of data
continues to increase endlessly. Astronomy is no exception. The Gaia
satellite is currently monitoring more than a billion stars with trillions
of observations. Transient surveys, such as Pan-STARRS, Skymapper
and the upcoming LSST, monitor billions of objects in the night sky
to detect stellar explosions. Such large numbers are difficult to
comprehend on a human level. Even the human population of Earth, currently
in excess of $7\times10^{9}$, is difficult to imagine. To modern,
massively parallel computers, the challenge is rather less daunting.
Commercial companies, especially in the financial sector, process
such data daily. There are thus natural links to astrophysics, where
the challenge is to understand the stars and galaxies we see, both
individually and as a collective. This is where population synthesis
modelling is such a useful tool. The general idea is to simulate a
population of stars on a computer, and compare simulated observables
to those we see. By varying the input parameters of the population
model to match the stars that are seen, we can pin down the underlying
astrophysics. Often, new physics can be ruled out because the numbers
fail to match, and uncertain physics can be far better constrained.

Population synthesis is particularly useful because astrophysics is
mostly not an experimental science. We want to answer fundamental
questions, such as how do stars function, how did they get there and
what are they going to do in the future. But we cannot \textendash{}
yet! \textendash{} experiment on stars. This is analogous to a dendrologist\index{dendrology}
examining the trees of a forest\footnote{Many thanks to Dr. Karl Kruszelnicki for the excellent analogy.}.
Some are small, some are large, some are green and some yellow, some
are in flower and some are dead. From this snapshot in time the dendrologist
must determine the entire biology and evolutionary history of trees:
a difficult task! Astronomers have been trying to do this for millennia
with the stars in the sky. Although we cannot experiment on stars
directly, through population synthesis we can experiment inside our
computers and statistically compare to our forest of stars.

First, we make a set of models of stars to attempt to match those
under investigation. Quantitative computer models of stars have been
constructed for at least fifty years. The current state of the art
is to combine one-dimensional stellar structure equations with parametrised
models of internal mixing and rotational deformation. Despite these
simplifications, such computer codes take hours to compute the evolution
of a single star from birth to death, even when details such as thermal
pulses and supernovae are ignored. Assuming the evolution can be calculated,
the second step is to turn such models into a stellar population.
For this, we need a description of the population's initial parameters
such as an initial mass function. Next, we must convert the models
into simulated observations. This is rarely a trivial task. Stellar
models predict physical parameters such as luminosity, effective temperature
and chemical abundances. Observations are of usually of stellar apparent magnitudes,
colours and spectral line widths, so conversion is required. Survey
selection biases must be taken into account when making the simulated
population. If such biases are simple \textendash{} say, a luminosity
cut off \textendash{} then this is relatively easy. In general, biases
are poorly known and sometimes impossible to quantify. Finally, our
population model should be compared to the observations in a quantitative
way to tell us which model is best and as much as possible about the
underlying astrophysics. Each of the above problems is its own challenge.
Population synthesis thus pushes the limits of both theoretical and
computational astrophysics.

\subsection{The Single and Binary Star Parameter Spaces}

\label{IzzardSubcec2-1}The stellar parameter space\index{stellar parameter space} is so large
that simulation of every star in the sky, in detail, is impossible.
We can, however, use our astrophysical knowledge to define a few key
initial parameters that describe the evolution of a star. Usually
these are, in order of importance, stellar mass $M$, metallicity\footnote{Metallicity, $Z$, is usually defined in stellar evolution as the
mass content of a star that is not hydrogen or helium. Observational
astronomy usually defined metallicity as $[\mathrm{Fe}/\mathrm{H}]=\log_{10}\left(N_{\mathrm{Fe}}/N_{\mathrm{Fe}\odot}\right)-\log_{10}\left(N_{\mathrm{H}}/N_{\mathrm{H}\odot}\right)$
where $N_{x}$ are the abundances, by number of particles, of species
$x$. Conversion between the two can be done e.g.~assuming a solar-scaled
$Z$.} $Z$ and rotation rate $v_{\mathrm{rot}}$. In the crudest approximation,
the rotation rate and metallicity are fixed, usually $Z=0.02$ ('solar'
metallicity) and $v_{\mathrm{rot}}=0$. This can be justified because
most stars have a metallicity roughly equal to that of the Sun and
most stars rotate slowly enough that their structure is not affected
substantially. It is thus possible to make a model population of $N$
stars by varying only their mass $M$. If each star takes $\Delta t=1\,\mathrm{h}$
to run on a modern CPU and we sample $M$ between, say, $0.8\mathrm{\,M_{\odot}}$
and $100\mathrm{\,M_{\odot}}$, we may need $N=100$ stars and the
simulation takes $N\times\Delta t=100\,\mathrm{hours}$. This is certainly
tractable.

Binary stars extend the parameter space by introducing a secondary
star of mass $M_{2}$, an orbital semi-major axis $a$ (or period
$P$, linked to $M_{1}$ and $M_{2}$ by Kepler's law),\index{binary orbit} an orbital
eccentricity $e$. Stellar spins are usually assumed to align with
the orbit \citep{1981A+A....99..126H} which is often assumed circular,
i.e.~$e=0$ \citep{2002MNRAS_329_897H}. However, even with $e=0$
and aligned inclinations, the single-star parameter space of dimension
1 ($M$) becomes a binary-star parameter space of dimension 3 ($M_{1}$,
$M_{2}$ and $a$), hence we have $N^{3}$ stellar models to run.
If $N=100$, we require $10^{6}$ stellar models which take about
$2\times10^{6}\,\mathrm{h}$ of CPU time. This is possible on a modern
computing cluster, and indeed has been done (albeit at lower resolution,
e.g.~\citealp{2017arXiv171002154E}), but we are at the current computational
limit and hence exploration of the model parameter space is restricted.

Single-star models contain many parameters which can vary, such as
the initial abundance mixture, the initial angular momentum profile,
the stellar mass loss prescription, angular momentum loss prescriptions
(e.g.~magnetic braking), uncertain nuclear reaction rates, models
of internal mixing, e.g.~convective mixing model (such as mixing length
theory), rotational mixing efficiencies and magnetic field evolution.
In binary stars we must include the strength of tidal interaction,
mass and angular momentum transfer efficiencies, mass and angular
momentum accretion efficiencies, an algorithm to describe when mass
transfer is stable, companion induced enhanced mass loss, gravitational
radiation angular momentum loss, nova explosions and interaction with
the ejected material, discs and associated jets, winds and orbital
interactions, and supernova kicks. We also require a model for common
envelope evolution and to allow for the possibility that stars merge\@.
Triple and higher stellar multiples also exist. To be stable, these
are hierarchical, so a triple can be modelled approximately as a close
binary inside a wide binary \citep{2016ComAC...3....6T}: it is still,
approximately, binary-star evolution.

Finally, when making a population,\index{initial binary distributions} each model parameter is either
fixed or has an initial distribution. The best-known example of such
is the initial mass function, $\xi\left(M\right)$, where $M$ is
the initial stellar mass. This is often described as a set of power
laws in $M$ (e.g.~\citealp{Kroupa1993,2001MNRAS.322..231K}). However,
even this formalism is subject to uncertainty, particularly in massive
stars ($M\gtrsim10{\,\mathrm{M}_{\odot}}$). In binaries, the secondary
mass, orbital separation (or period) and eccentricity all have their
own input distributions, and a mass-dependent binary fraction is
often employed \citep{2010ApJS..190....1R,2012Sci...337..444S,2013arXiv1303.3028D,2017ApJS..230...15M,2017ApJ...836..139F}.

So, in general, modelling involves a large number of input parameters
$N_{\mathrm{param}}$. In our \emph{binary\_c} single and binary-star
population synthesis code there are currently 238 model parameters
for each binary star, not including the many parameters which are
compile-time constants and parameters that form the initial distributions.
Thus $N_{\mathrm{param}}\approx1000$ is not unreasonable.

Which parameters are important depends very much on which stars are
being modelled and compared to. For example, changing mass loss rates
on the red giant branch will not affect main sequence lifetimes, so
if one is measuring the time spent on the main sequence this parameter
is unimportant. However, if one is measuring the number of main sequence
stars which have accreted material from chemically peculiar red giants
then the giant's mass loss rate is very important. There is currently
no easy way to determine \emph{a priori} which parameters are important
to make which stars: the experience of the astrophysicist is crucial
in deciding. This task is, however, vital. If $N_{\mathrm{param}}$
can be reduced from $1000$ to just a few, the problem is greatly
reduced in complexity and may become tractable even if the model run
time, $\Delta t$, is long. In general, however, $N_{\mathrm{param}}$
is so large that it is necessary to reduce $\Delta t$ by many orders
of magnitude as described next.

\subsection{Detailed and Synthetic Stellar Models}

\label{IzzardSubcec2-2}Full stellar evolution models of some
stars, such as those in the thermally-pulsing asymptotic giant branch
(TPAGB) or those undergoing rapid mass transfer, are far more costly
to create than the $\Delta t\sim1\,\mathrm{h}$ suggested previously.
$\Delta t$ in these cases could be days, weeks or longer. It is impossible
to make population models containing such stars covering any significant
parameter space at reasonable resolution simply because of time constraints.
Students have PhDs to finish and even tenured astrophysicists retire
eventually!

One solution is to make a grid of models. The grid can be interpolated
(e.g.~\citealp{2002NewA....7...55D,2011A&A...530A.116B,Kruckow2017})
or fitted to functions (e.g.\emph{~SSE }and \emph{BSE} of \citealp{1989ApJ...347..998E,1997MNRAS.291..732T,2000MNRAS.315..543H,2002MNRAS_329_897H})
to reduce the run time \textendash{} often to less than $1\,\mathrm{s}$.
With $\Delta t=1\,\mathrm{s}$, a typical binary grid takes about
$10^{6}\,\mathrm{s}$, or about one day on a 16-core PC. Stellar codes
which employ such techniques are referred to as \emph{synthetic }stellar
evolution codes, as opposed to their \emph{detailed} full stellar
evolution cousins. In the case of fitted formulae, internal stellar
structure is sacrificed to gain a factor of at least $10^{6}$ in run
time. Grid lookup codes can retain internal structure which is useful
when calculating, e.g., the stellar binding energy, at the cost of
computational storage space.\index{synthetic stellar models}

Our \emph{binary\_c}\index{binary\_c} code \citep{Izzard_et_al_2003b_AGBs,2006A&A...460..565I,2009A&A...508.1359I,2017arXiv170905237I}
uses the a \emph{C} version of the \emph{BSE} library for most of
its stellar structure calculations and nucleosynthesis from various
sources, including \citet{Karakas2002} and \citet{Karakas2009update}
during the TPAGB, and supernova yields from massive stars \citep{1995ApJS..101..181W,2004ApJ...608..405C}.
It runs at least $10^{6}$ times faster than the full stellar evolution
codes on which its is based, but still provides useful estimates of
stellar luminosity, mass, core mass, radius and core radius, and chemistry
as a function of time for the entire evolution of the star. Binary-star
interaction is included mostly according to \citet{2002MNRAS_329_897H}
but with improvements to mass transfer by Roche-lobe overflow \citep{2014A&A...563A..83C},
Wind-Roche-Lobe-overflow \citep{2013A&A...552A..26A,2015A&A...581A..62A}
, tides \citep{2013A&A...550A.100S}, rejuvenation \citep{2013ApJ...764..166D,2014ApJ...780..117S}
and supernovae \citep{2017MNRAS.469.2151B,2017A&A...606A..14B}. Binary-star
nucleosynthesis includes accretion and thermohaline mixing \citep{2007A&A...464L..57S,2017arXiv170905237I}
as well as explosions such as thermonuclear novae \citep{Jose1998}
and Ia supernovae \citep{2014A&A...563A..83C}. Modern detailed binary
star codes, including \emph{MESA} \citep{2011ApJS..192....3P,2015ApJS..220...15P},
implement similar binary interaction physics to \emph{BSE} and \emph{binary\_c},
and take far longer to run. If one is interested in the effects of
binaries on a stellar population, rather than the precise details
of stellar structure, it makes no sense to throw endless CPU cycles
at the problem.

Synthetic codes have another major advantage over detailed codes,
that of stability. While detailed stellar evolution codes habitually
stop because of physical or numerical problems (they 'crash'), synthetic
codes rarely suffer such problems. Thus, if a task does not require
full, detailed stellar evolutionary calculations, a synthetic code
is often a good choice for both speed and reliability. Often the observations
to which the models are being compared have uncertainties that do
not justify more accurate, and costly, modelling.

\section{Stellar Accountancy}

\label{IzzardSec3}Our aim is to model a population of stars which
matches the stars we see now.\index{stellar accountancy} These stars were born in the past, at
rates given by a star formation rate $S(t)$, where $t$ is time since
the Big Bang. Stars in our model are labelled $i$ and each has a
probability of existing, $\psi_{i}$, normalised such that $\sum_{i}\psi_{i}=1$.
Our model then predicts the number of stars of type $j$ that exist
now, when $t=t_{\mathrm{max}}$, is, \index{stellar number counting}
\begin{eqnarray}
  N_{j} & = & \sum_{t=0}^{t_{\mathrm{max}}}S\left(t,\,\mathbf{Z}\right)\sum_{i}\psi_{i}\left(M,\dots,\mathbf{Z}\right)\sum_{\tau=0}^{t_{\mathrm{max}}-\tau}\delta_{j}\left(t-\tau,\,\mathbf{X},\,\mathbf{Z}\right)\,\delta\tau\,,
\end{eqnarray}
\label{IzzardEq:number-of-stars}
where $\tau$ is the age of each star, $\delta\tau$ is the timestep
of the model and $\delta_{j}$ is a function that is $1$ if the star
is of the type $j$ and $0$ otherwise. 
\begin{figure}
\begin{centering}
\hspace{-0cm}%
\noindent\begin{minipage}[t]{1\columnwidth}%
\begin{center}
\begin{tabular}{cc}
  \textbf{a)}~~\includegraphics[width=6cm]{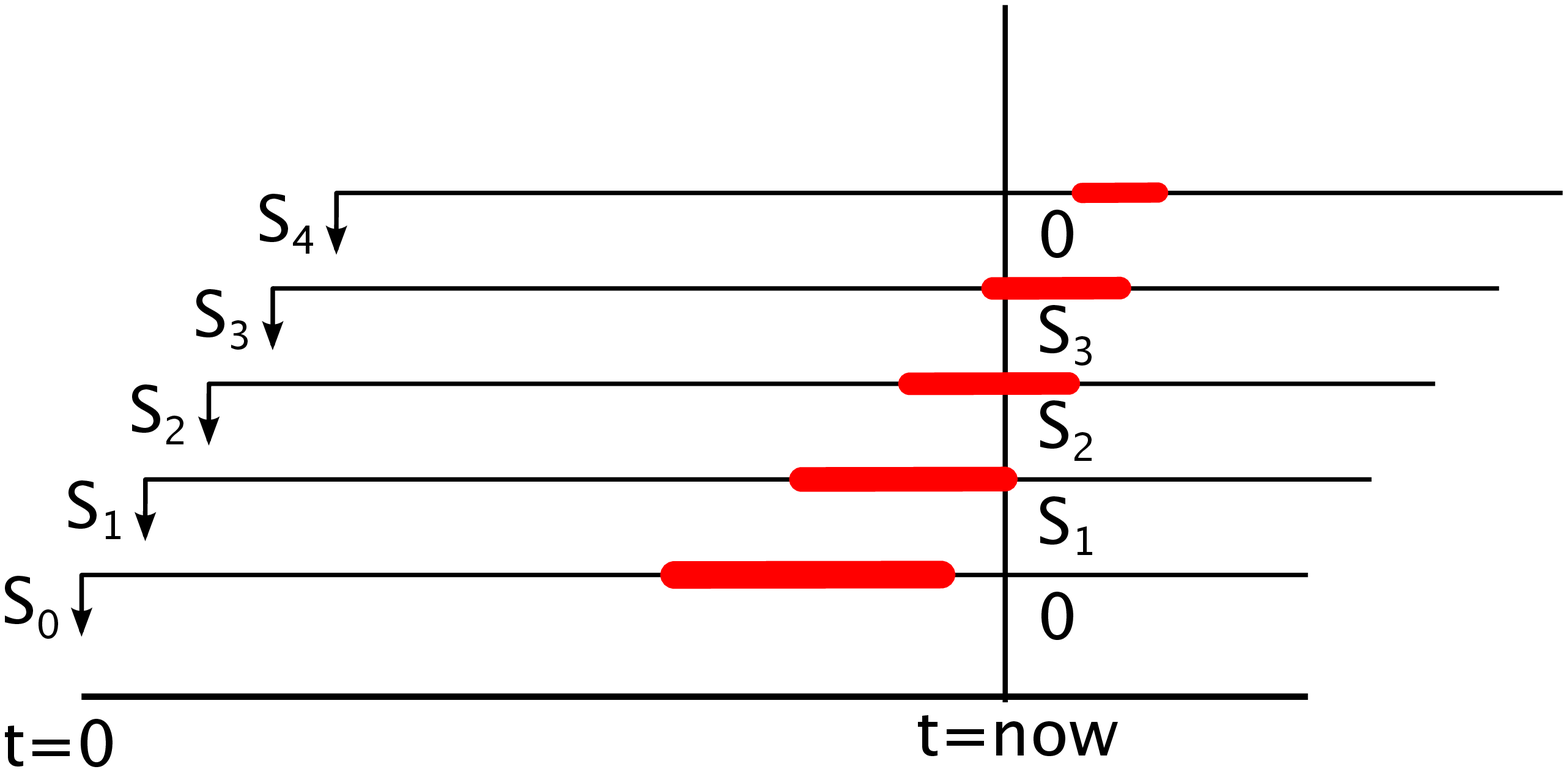} &
  \textbf{\textbf{\hspace{-0.7cm}}b)~~}\includegraphics[width=6cm]{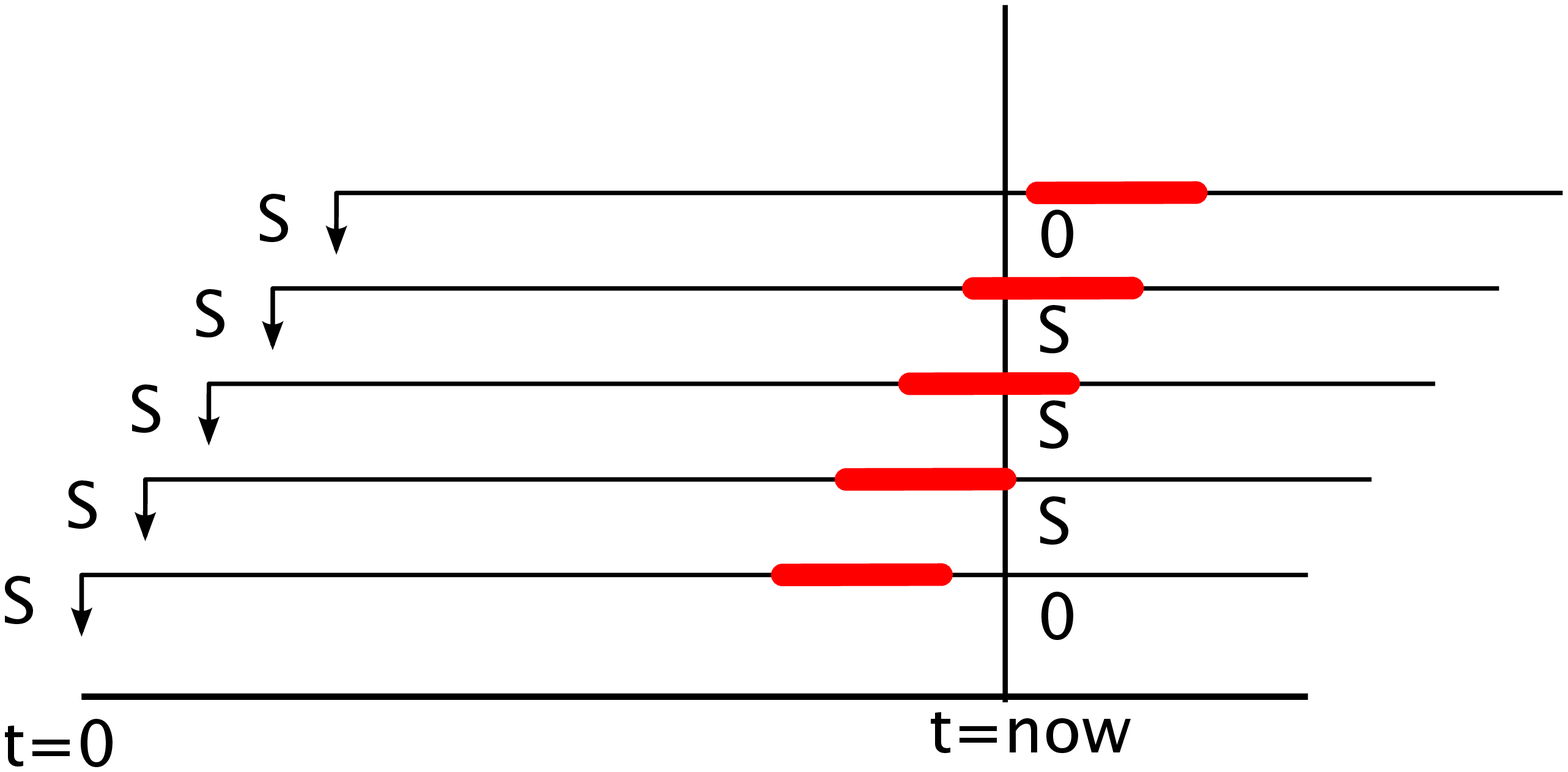} 
  \tabularnewline
  \multicolumn{2}{c}{
    \textbf{\textbf{\hspace{-0cm}}c)~~}\includegraphics[width=5cm]{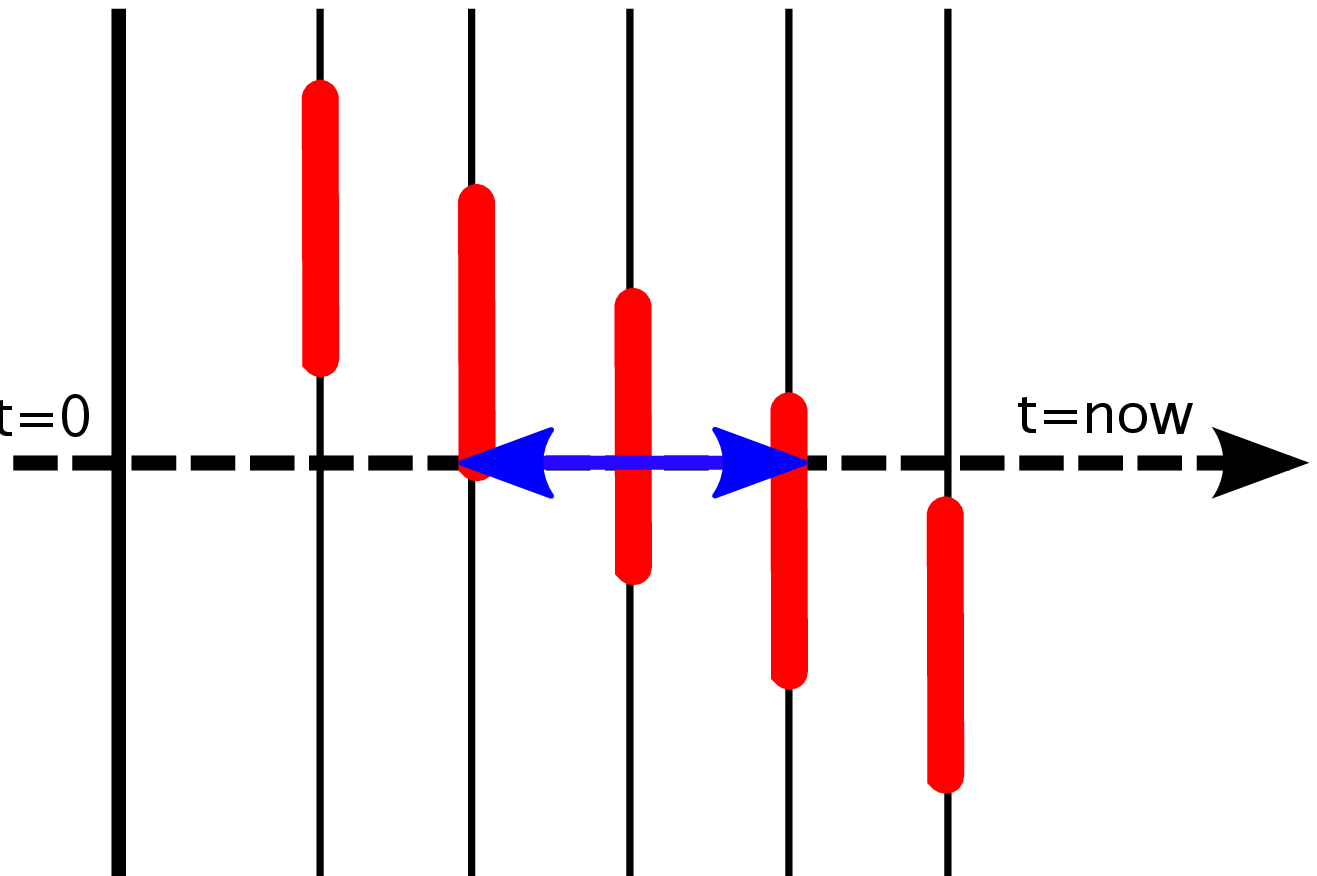}
  }
    \tabularnewline
\end{tabular}
\par\end{center}%
\end{minipage}
\par\end{centering}
\caption[Simplified population synthesis modelling]{
  \label{IzzardFig:1-convolution}
  Graphical representation of population
synthesis modelling. Time runs left to right from the Big Bang ($t=0$)
until now in finite time steps, $\Delta t$. Stars formed at $0$
to $4\Delta t$ are shown, with corresponding star formation rates
$S_{0},$ $S_{1}$, \ldots{}, but in a realistic simulation there
may be many thousands of time steps. Stars evolve from left to right
along horizontal lines which, when they are in the phase of interest
that is to be counted, they are thick and red (equivalent to $\delta_{j}=1$
in Eq.~(\ref{IzzardEq:number-of-stars}), $\delta_{j}=0$ otherwise). Stars
are counted when the phase of interest intersects the vertical line
which represents now. \textbf{a)} In an ideal population model, at
each time step the star formation rate varies ($S_{0}\protect\neq S_{1}\protect\neq S_{2}\dots$)
and the properties of the stars also vary (indicated by the change
in colour and length of time spent in the phase of interest). \textbf{b)
}In a simplified model, the stars all have the same properties (e.g.~the
same metallicity) and hence length of time in the phase of interest.
The star formation rate is constant, i.e.~$S=S_{0}=S_{1}=S_{2}\dots$.
\textbf{c) }Instead of performing an expensive convolution over many
models, the problem is rotated by $90^{\circ}$ and population statistics
are calculated from a single model run represented by the horizontal
dashed black line. The region denoted by the blue arrow is the phase
of interest and the evolutionary model set has to be calculated only
once, so the calculation is relatively cheap.
}
\end{figure}
The vector $\mathbf{Z}$ represents model input parameters, such as
metallicity, and input distributions which can change with the time
of birth of the stars. The vector $\mathbf{X}$ denotes the many parameters\footnote{Some parameters exist in both $\mathbf{X}$ and $\mathbf{Z}$ in our
rather crude definition.} which affect stellar evolution, such as metallicity, rotation rate,
mass-loss prescription and mixing-length parameter. The function $\delta_{j}$
factors in stellar evolution, so is a function of $\mathbf{X}$, $\mathbf{Z}$
and $\tau$. The probability, $\psi_{i}$, is a function of at least
mass, perhaps also metallicity (hence $\mathbf{Z}$) and, for binary
stars, the companion mass and orbital properties also. If the function
$\psi_{i}$ has $n$ parameters, the sum $\sum_{i}$ is implied over
all of them. The large number of elements of this sum illustrates
the parameter space problem we face even if $\mathbf{X}$ and $\mathbf{Z}$
are fixed.

In its most general form, Eq.~(\ref{IzzardEq:number-of-stars}) is a nasty
convolution problem of the type faced by galactic chemical evolution
models \citep{2003PASA...20..401G}. The problem can be simplified
by fixing the stellar physics (i.e.~$\mathbf{X}=\mathrm{constant}$
and $\mathbf{Z}=\mathrm{constant}$), the input distributions and
assuming a constant star formation rate, e.g.\footnote{Formally $S=T^{-1}$ where $T$ is the unit of time in which $\delta\tau$
is measured. A time-dependent star formation rate can also be implemented
as $S\left(\tau\right)$, provided $\mathbf{X}$ is fixed, but we
limit ourselves to the simpler case of constant star formation rate.}~$S=1$. The sum reduces to, as shown in Fig.~\ref{IzzardFig:1-convolution},
\vspace{-2mm}
\begin{eqnarray} N_{j} & = & \sum_{i}\psi_{i}\left(M,\dots\right)\sum_{\tau=0}^{t_{\mathrm{max}}}\delta_{j}\left(\tau\right)\,\delta\tau\,.\label{IzzardEq:number-of-stars-reduced}
\end{eqnarray}
The parameter space wrapped into $\psi_{j}$ is still potentially
very large, but the number of models to be calculated is greatly reduced
because $\mathbf{X}$ and $\mathbf{Z}$ are fixed. We can also reduce
the size of the sum over time by fixing the time step, $\delta\tau$,
and allowing $\delta_{j}$ to take values between $0$ and $1$ corresponding
to partial occupancy in phase $j$ during each time step. This is
useful when calculating tables of model results with a fixed time
resolution, e.g.~for later use in a galactic chemical evolution context.
The number of stars in a galaxy can be predicted by then scaling $S$
appropriately. 

All the above is for just one type of star, denoted by $j$. In reality
there are many different types of star and hence many different $N_{j}$.
Our binary-star population synthesis code \emph{binary\_c} has more
than 200 such types of star, both single and binary. It is also possible
to count stars in bins, so that individual $N_{j}$ can be combined
to create distributions. Differential measurements, such as number
ratios, are more useful than simple predictions of the number of different
types of stars in a population. One would hope that errors in approximations,
such as our assumption that the star formation rate is constant, at
least partly cancel out.

Further, the definition can be extended to count, say, stellar mass
ejecta as we show in the Section~\ref{IzzardSec4-Slow-and-fast}.
We set $\delta_{j}=\dot{M}_{j}\left(\tau\right)$, the rate of mass
lost from a system as isotope $j$ as a function of stellar age, $\tau$.
The normalised mass ejected as isotope $j$ from a population of single
stars is defined as,
\begin{eqnarray}
y_{j,\mathrm{sin}} & = & \frac{\sum_{i=M}\psi_{i}\sum_{\tau=0}^{t_{\mathrm{max}}}\dot{M}_{j}\left(\tau\right)\,\delta\tau}{\sum_{i=M}\psi_{i}M}\,,\label{Izzard:Eq-single-star-ejecta}
\end{eqnarray}
and from a population of binary stars as,\index{mass ejecta}\index{chemical yields}
\begin{eqnarray}
y_{j,\mathrm{bin}} & = & \frac{\sum_{i=M_{1},M_{2},a}\psi_{i}\sum_{\tau=0}^{t_{\mathrm{max}}}\dot{M}_{j}\left(\tau\right)\,\delta\tau}{\sum_{i=M_{1},M_{2},a}\psi_{i}\left(M_{1}+M_{2}\right)}\,,\label{IzzardEq-binary-star-ejecta}
\end{eqnarray}
where $M$ is the initial mass of a single star and $M_{1}$, $M_{2}$
and $a$ are the initial primary mass, initial secondary mass and
initial separation of a binary. The sums in the denominators are the
total mass formed in stars in each population. This must be included
to compare single to binary stars because binary systems have initially
more mass, on average, than single stars. Note that $y_{j}$ are dimensionless.

\section{Slow and Fast Parameters}

\label{IzzardSec4-Slow-and-fast}Eq.~(\ref{IzzardEq:number-of-stars-reduced})
contains two main sets of parameters. The \emph{fast} parameters are
those that affect the input distributions, $\psi_{i}$. These include
the slope and shape of the initial mass function, the secondary mass
(or mass ratio) distribution and the initial orbital period distribution.
These are called \emph{fast} because changing $\psi_{i}$ does not
require a recalculation of the stellar evolution model (which is embedded
in $\delta_{j}$). Indeed, at any timestep in a stellar population
calculation, as many different $\psi_{i}$ as desired can be calculated.

The \emph{slow} parameters, on the other hand, require a complete
recalculation of the stellar structure and evolution at every timestep.
These are the $\mathbf{X}$ of Eq.~(\ref{IzzardEq:number-of-stars}). The
calculation process is thus, in terms of computer speed, slow. Such
parameters include the metallicity, mixing parameters, mass loss prescriptions
and nuclear reaction rates. When stellar model grids are calculated,
most of the slow parameters are fixed. Only one or a few parameters
are then varied, such as the metallicity. Many of the constant parameters
may not matter greatly to stellar evolution, but this often depends
on the type of star under consideration, so the precise definition
of our $j$ and hence $\delta_{j}$, and this dependence is often
not known \emph{a priori}.

In the following we demonstrate the dependence of a stellar population
statistic, the amount of mass ejected from a stellar population as
a particular isotope, on various model input parameters\footnote{These models are from an old version of \emph{binary\_c} first presented
in the first author's PhD dissertation but demonstrate the principle
under consideration accurately enough.} using Eqs.~\ref{Izzard:Eq-single-star-ejecta} and~\ref{IzzardEq-binary-star-ejecta}.
The fast parameters affect the initial distributions, $\psi_{i}$,
while the slow parameters affect the $\dot{M}_{j}\left(\tau\right)$,
i.e.~when and how much mass is ejected as isotope $j$ as a function
of time. The initial parameters, $i$, are sampled on logarithmic
grids, with single star masses in the range $0.1\leq M/\mathrm{M_{\odot}}\leq80$
and $n_{M}=1000$ stars, binary component masses $M_{1}$ and $M_{2}$
are treated identically to $M$ but with $0.1\leq M_{2}\leq M_{1}$,
$n_{M1}=100$ and $n_{M2}=100$, and separations in the range $3\leq a/\mathrm{R_{\odot}}\leq10^{4}$,
also with $n_{a}=100$ (so the total number of binary systems in each
grid is $n_{M1}\times n_{M2}\times n_{a}=10^{6}$). At this resolution,
errors in the relative mass ejected caused by a limited resolution
and timestep are at most $2\%$.

\subsection{Fast Parameters}

\label{IzzardSubsec4-1}\index{fast parameters}In our example we have three distributions
that define the fast parameters. These are the single or primary star
initial mass distribution, the initial secondary mass distribution
and the initial separation distribution. The single or primary star
mass distribution is also known as the initial mass function. Our
standard model is the \citet{Kroupa1993} initial mass function and
we consider two alternatives from \citet{2003PASP..115..763C} and
\citet{1955ApJ...121..161S}. Applying each gives us a range of
uncertainty in stellar ejecta, $y_{j}$, as shown in Table~\ref{IzzardTable:1-fast-parameter-table}.

We apply a similar process to the binary-star primary mass function
which is assumed to be identical to that of single stars. The results
are similar to those in single stars, but the impact of binary stars
is seen in the reduced nitrogen and barium ejecta, because mass transfer
truncates the evolution of thermally pulsing asymptotic giant branch
stars, and in an increase in iron because of type Ia supernovae.

Our default distribution of secondary star masses is assumed to be
flat in $q=M_{2}/M_{1}$, i.e.~$\psi_{q}\propto q^{0}$, between $q=q_{\mathrm{min}}=0.1\,\mathrm{M}_{\odot}/M_{1}$
and $1$. We also consider distributions with $\psi_{q}\propto q^{0.5}$
(cf.~\citealp{1980ApJ...242.1063G}) and $\psi_{q}\propto q^{-1.5}$,
for illustration. In general, the changes to the mass ejecta, $y_{j}$,
are smaller than changes caused by variation in the initial or primary
star mass distribution.

Finally, we consider the distribution of binary orbits. This is set
either by a distribution of orbital semi-major axes, $a$, or orbital
periods, $P$, where both are related by Kepler's law (we assume circular
orbits). Our default distribution is $\psi_{a}\propto a^{-1}$, equivalent
to \citet{1924PTarO..25f...1O}. We also consider $\psi_{a}\propto a^{1}$
, $\psi_{a}\propto a^{-2}$ and combination of $\psi_{a}\propto a^{-0.7}$
and $10\leq a/\mathrm{R_{\odot}}\leq1.3\times10^{3}$ to match \citet{2003ApJ...591..397G}.
Table~\ref{IzzardTable:1-fast-parameter-table} shows that, as with
the secondary-star mass distribution, reasonable changes in the separation
distribution matter less than reasonable changes in the initial/primary
star mass distribution.

Please note that the numbers calculated here are only a demonstration.
A proper exploration of this problem would consider the more recent
distributions of, e.g., \citet{2017ApJS..230...15M} with their error
bars. Such an exploration is unfortunately beyond the scope of this
work.

\begin{table}[h!]
\begin{minipage}{352pt}
  
    \addtolength\tabcolsep{10pt}
    \caption[Ejecta from a stellar population as a function of fast parameters.]{Masses of various isotopes
ejected from a stellar population relative to the mass born into stars
when reasonably varying \emph{fast} parameters, i.e.~the initial
mass function in single stars, and in binary stars the primary star
mass function, secondary star mass distribution and initial orbital
separation within reasonable limits. These parameters are \emph{fast}
because they require only one run of the stellar evolution models.}
\label{IzzardTable:1-fast-parameter-table}

    \begin{tabular}{|c|c|c|c|c|}
\hline  \hline
\multirow{2}{*}{Isotope} & Single stars & \multicolumn{3}{c|}{Binary stars}\tabularnewline
 & Vary $M$ & Vary $M_{1}$ & Vary $M_{2}$ & Vary $a$\tabularnewline
\hline 
$^{1}\mathrm{H}\times10$ & $1.9-3.7$ & $1.9-3.6$ & $1.7-2.2$ & $1.6-2.6$\tabularnewline

$^{12}\mathrm{C}\times10^{3}$ & $4.6-15$ & $5.7-18$ & $3.8-6.1$ & $4.0-6.2$\tabularnewline

$^{14}\mathrm{N}\times10^{4}$ & $9.6-24$ & $8.1-19$ & $7.6-9.8$ & $7.9-9.5$\tabularnewline

$^{16}\mathrm{O}\times10^{3}$ & $8.4-30$ & $9.6-32$ & $7.2-10$ & $7.1-11.5$\tabularnewline

$^{56}\mathrm{Fe}\times10^{3}$ & $1.0-3.0$ & $2.5-5.4$ & $1.3-2.7$ & $0.9-3.4$\tabularnewline

$^{65}\mathrm{Cu}\times10^{6}$ & $10-32$ & $8.8-28$ & $7.0-9.2$ & $8.0-12$\tabularnewline

$\mathrm{Ba}\times10^{9}$ & $8.1-16$ & $6.5-13$ & $6.0-7.8$ & $6.2-9.4$\tabularnewline
\hline  \hline
\end{tabular}
\par
\end{minipage}
\end{table}

\subsection{Slow Parameters}

\label{Izzard:Subsec-slow-parameters}\index{slow parameters}There are many \emph{slow}
model parameters that alter the chemical ejecta from a population
of stars. We limit our investigation to a population of 100\% binary
stars with fixed initial distributions of masses and orbital separations.
Primary masses $M_{1}$ are distributed according to \citet{Kroupa1993}
between $0.1$ and $80\mathrm{\,M_{\odot}}$, secondary mass ratios
$q$ follow a flat distribution between $0.1\mathrm{\,M_{\odot}/M_{1}}$
and $1$, and separations, $a$, have a constant distribution in $\log a$
between $3$ and $10^{4}\mathrm{\,R}_{\odot}$.

The parameters we vary are metallicity in the range $10^{-4}\leq Z\leq0.02$
(default is $0.02$), binary eccentricity\textbf{ }$0\leq e\leq1$
(default is $0$), red giant branch mass loss parameter $0.25\leq\eta\leq0.75$
(default is $0.5$ using the \citealp{1975psae.book..229R} formalism
as defined in \citealp{2000MNRAS.315..543H}'s Eq.~{[}106{]}), $^{13}\mathrm{C}$
pocket efficiency $0.01\leq\xi_{13}\leq2.0$ (default is $1$, \citealp{2006A&A...460..565I}),
wind loss rate during a Wolf-Rayet phase multiplier $0.1\leq f_{\mathrm{WR}}\leq10$
(default is $1$, cf.~\citealp{2000MNRAS.315..543H}'s section~7.1),
an Eddington limit for accretion multiplied by $1$ or $10^{6}$ (default
is $10^{6}$ i.e.~no limit, \citealp{2002MNRAS_329_897H}'s Eq.~{[}67{]}),
a supernova kick velocity dispersion $0\leq\sigma_{\mathrm{SN}}/\mathrm{km}\,\mathrm{s}^{-1}\leq400$
(default is $190\,\mathrm{km}\,\mathrm{s}^{-1}$, \citealp{Hansen1997}),
companion reinforced attrition process \citep{1988MNRAS.231..823T}
parameter $0\leq B\leq10^{4}$ (default is $0$, i.e.~no reinforcement),
common envelope parameter $\alpha_{\mathrm{CE}}=3$ \citep{2002MNRAS_329_897H}
and third dredge up parameters $-0.1\leq\Delta M_{\mathrm{c},\mathrm{min}}\leq0$
(default $-0.07\mathrm{\,M_{\odot}}$) and $0\leq\lambda_{\mathrm{min}}\leq1$
(default $0.8-37.5Z$ , \citealp{Izzard_et_al_2003b_AGBs,Binary_Origin_low_L_C_Stars}).

The normalised ejected mass of each isotope, $y_{j}$, is expanded
as a Taylor series in the slow parameters labelled $k$ with values
$x_{k}$, 
\begin{eqnarray}
y_{j}\left(x_{k}\right) & = & y_{j0}+\frac{dy_{j}}{dx_{k}}\left(x_{k}-x_{k0}\right)+\dots\approx y_{j0}+\left(1+T_{jk}\frac{\delta x_{k}}{\Delta x_{k}}\right)\,,\label{Izzardeq:taylor}
\end{eqnarray}
where $x_{k0}$ are the default values which one may consider best
estimates, $y_{j0}$ are the ejecta corresponding to the default parameter
set $x_{k0}$, $\delta x_{k}=x_{k}-x_{k0}$, and $x_{k}$ varies from
$x_{k,\mathrm{min}}$ to $x_{k,\mathrm{min}}+\Delta x_{k}$. The matrix
of derivatives,
\begin{eqnarray}
T_{jk} & = & \frac{d\left(y_{j}/y_{j0}\right)}{d\left(x_{k}/\Delta x_{k}\right)}\,,\label{Izzard:eq:Tjk}
\end{eqnarray}
provides a relatively unbiased comparison of parameters because it
measures the rate of change of dimensionless and normalised numbers,
$y_{j}/y_{j0}$ and $x_{k}/\Delta x_{k}$. A larger absolute value
of $T_{jk}$ thus reflects greater sensitivity of amount of isotope
$j$ ejected to the parameter $k$. One drawback of this approach
is that it only considers the linear terms in the Taylor expansion:
it is possible that, at second or higher order, the effects of changing
two parameters cancel each other out. Such considerations are beyond
the scope of this work.

Table~\ref{IzzardTable:2-slow-parameters} shows how the $T_{jk}$
vary with each model parameter. Absolute values above $0.1$ are marked
in bold/red. These are the parameters to which the chemical ejecta
are most sensitive. As an example, the ejecta of most isotopes vary
little with the $^{13}\mathrm{C}$ $s$-process pocket efficiency.
The exception is barium, an $s$-process element, which is very sensitive
to the efficiency of $s$-processing, as one would expect. While this
is a trivial example, it is perhaps less obvious that the ejection
of $^{16}\mathrm{O}$ depends ten times more weakly on the common
envelope parameter, $\alpha_{\mathrm{CE}}$, than does $^{14}\mathrm{N}$.

\begin{table}[h!]
\begin{minipage}{360pt}
\addtolength\tabcolsep{-5.0pt}
\caption[Ejecta from a stellar population as a function of slow parameters.]{Ejecta dependence on various
model parameters as measured by $T_{jk}$, the magnitude of which
is an attempt to measure the dependence on each model parameter $k$
of the relative mass ejected from a population of stars as isotope
$j$ (Eq.~\ref{Izzard:eq:Tjk}, symbols are defined in the main text).
$x^{(y)}$ means $x\times 10^{y}$.
The numbers highlighted in red/bold have an absolute value exceeding
$0.1$ and are the most sensitive to their parameter.

}
\label{IzzardTable:2-slow-parameters} 

{\small{}\hspace{-0.35cm}}%
\noindent\begin{minipage}[t]{1\columnwidth}%
{\small{}}%
\begin{tabular}{|r|r|r|r|r|r|r|r|r|r|r|}
\hline \hline
\multirow{2}{*}{{Isotope}} & \multicolumn{10}{c|}{{Model Parameter}}\tabularnewline
 & {$Z$} & {$\xi_{\mathrm{C}13}$} & {$f_{\mathrm{WR}}$} & {$\alpha_{\mathrm{CE}}$} & {$\Delta M_{\mathrm{c},\mathrm{min}}$} & {$e$} & {$f_{\mathrm{Edd}}$} & {$\eta_{\mathrm{GB}}$} & {$\lambda_{\mathrm{min}}$} & {$\sigma_{\mathrm{SN}}$}\tabularnewline

\hline 
{$^{1}\mathrm{H}$} & \textcolor{red}{$\mathbf{-1.9^{(-1)}}$} & {$1.2^{(-4)}$} & {$-3.6^{(-3)}$} & {$-1.6^{(-2)}$} & {$5.2^{(-3)}$} & {$4.5^{(-2)}$} & {$-3.5^{(-4)}$} & {$8.2^{(-3)}$} & {$-9.9^{(-3)}$} & {$-5.3^{(-3)}$}\tabularnewline

{$^{12}\mathrm{C}$} & {$6.0^{(-2)}$} & {$-1.8^{(-3)}$} & \textcolor{red}{$\mathbf{1.0^{(-1)}}$} & \textcolor{red}{$\mathbf{1.1^{(-1)}}$} & {$-7.1^{(-2)}$} & \textcolor{red}{$\mathbf{-1.6^{(-1)}}$} & {$-1.8^{(-2)}$} & {$-5.0^{(-2)}$} & \textcolor{red}{$\mathbf{1.6^{(-1)}}$} & {$4.6^{(-2)}$}\tabularnewline
{$^{14}\mathrm{N}$} & \textcolor{red}{$\mathbf{-6.5^{(-1)}}$} & {$-1.3^{(-4)}$} & {$-5.9^{(-3)}$} & \textcolor{red}{$\mathbf{-4.1^{(-1)}}$} & {$-3.8^{(-3)}$} & {$9.4^{(-2)}$} & {$-2.7^{(-5)}$} & {$-1.9^{(-2)}$} & {$1.8^{(-2)}$} & {$2.7^{(-2)}$}\tabularnewline
{$^{16}\mathrm{O}$} & {$8.8^{(-2)}$} & {$-2.6^{(-3)}$} & \textcolor{red}{$\mathbf{-4.9^{(-1)}}$} & {$-2.3^{(-2)}$} & {$2.0^{(-3)}$} & {$-5.3^{(-2)}$} & {$1.2^{(-3)}$} & {$-1.0^{(-2)}$} & {$-4.0^{(-4)}$} & \textcolor{red}{$1.0^{(-1)}$}\tabularnewline
{$^{56}\mathrm{Fe}$} & \textcolor{red}{$\mathbf{9.6^{(-1)}}$} & {$-3.9^{(-4)}$} & \textcolor{red}{$\mathbf{-1.8^{(-1)}}$} & \textcolor{red}{$\mathbf{2.3^{(-1)}}$} & {$1.8^{(-3)}$} & \textcolor{red}{$\mathbf{-2.5^{(-1)}}$} & {$5.8^{(-2)}$} & {$-9.3^{(-3)}$} & {$-2.5^{(-3)}$} & {$2.5^{(-2)}$}\tabularnewline
{$^{65}\mathrm{Cu}$} & \textcolor{red}{$\mathbf{-7.2^{(-1)}}$} & {$-5.8^{(-4)}$} & {$-7.5^{(-2)}$} & \textcolor{red}{$\mathbf{-3.1^{(-1)}}$} & {$1.0^{(-3)}$} & {$7.8^{(-2)}$} & {$1.1^{(-3)}$} & {$-2.0^{(-2)}$} & {$1.4^{(-3)}$} & {$7.0^{(-2)}$}\tabularnewline
{$\mathrm{Ba}$} & \textcolor{red}{$\mathbf{-2.2}$} & \textcolor{red}{$\bm{1.3}$} & {$2.5^{(-2)}$} & \textcolor{red}{$\mathbf{-2.6^{(-1)}}$} & {$-9.1^{(-2)}$} & \textcolor{red}{$\mathbf{1.1^{(-1)}}$} & {$-2.2^{(-3)}$} & {$-7.1^{(-2)}$} & \textcolor{red}{$\mathbf{2.0^{(-1)}}$} & {$3.1^{(-3)}$}\tabularnewline
\hline \hline
\end{tabular}
\end{minipage}
\end{minipage}
\end{table}

\section{Matching Models to Observations, and Models to Models}

\label{IzzardSec5}\index{Bayesian methods}\index{Matching observations to models}
The general problem of matching model results,
with their distributions of input parameters, to many observations
is one that is, in every sense of the word, non-trivial. The size
of the parameter space is daunting. Efforts like those mentioned in
Sec.~\ref{Izzard:Subsec-slow-parameters}
reduce the number of parameters to only those that matter for a given
observational data set of a particular type of star. Perhaps the greatest
problem is in how to match all stars from as many data sets as possible
simultaneously, thus provide the best possible constraints. The solution
to this problem is beyond our current reach. 

Frameworks to match a grid of models to an individual observed star
are relatively well developed. A good example is the Bayesian algorithm
\emph{BONNSAI} \citep{2014A&A...570A..66S,2017A&A...598A..60S}. Given,
say, the effective temperature, luminosity and surface abundances
of the star, all the models in a particular grid are compared to it
to determine \textendash{} in an automated way \textendash{} which
fits best. Bayesian techniques also naturally give posterior distributions
of parameters, hence associated error bars. \emph{BONNSAI} also tests
that model resolution is sufficient to believe the best fit. Other,
related techniques such as Markov-chain Monte Carlo (MCMC) methods
are starting to be employed by the community, particularly in the
hunt for the progenitors of gravitational wave sources \citep[e.g.][]{2010CQGra..27k4007M}.
The next big step will be to use such techniques in general, e.g.~to
match the data of \emph{Gaia}, to whole population models.

When matching a large number of stars to population models, the selection
effects of the survey in question should also be taken into account.
All surveys, even those that are 'complete', have some kind of
selection bias. In many cases, a simple cut in say magnitude or surface
gravity, which effectively selects stars in a particular phase of
evolution, is sufficient and easy to fit into a Bayesian analysis
(e.g.~\citealp{2014ApJ...787..110C,2017arXiv170905237I}). Modern
surveys often have a well described 'selection function' which
describes in a probabilistic way the various biases inherent to their
sample, e.g.~the \emph{Gaia-ESO} survey \citep{2016MNRAS.460.1131S}.
However, older data often has poorly defined selection effects and,
in some cases, data is selected 'by eye' which renders its modelling
essentially impossible.

Finally, our computation models must be tested and verified. One way
to do this is to compare the models of one group against those of
another to test for consistency. In the case that both groups use
identical input physics, the results should be the same. The \emph{POPCORN}
collaboration tried to do this, pitting \emph{binary\_c} against the
\emph{SeBa},\emph{ Startrack }and \emph{Brussels }population synthesis
codes \citep[and references therein]{2014A&A...562A..14T}. The four
codes were, for the most part, consistent, which at least suggests
we are getting the numerics right. Whether the physics they employed
is true to life is another question which \emph{POPCORN} did not try
to answer.

\section{Headline News in Population Synthesis}

\label{IzzardSec6}
\begin{figure}

\noindent\begin{minipage}[t]{1\columnwidth}%
\includegraphics[angle=270,width=0.5\paperwidth]{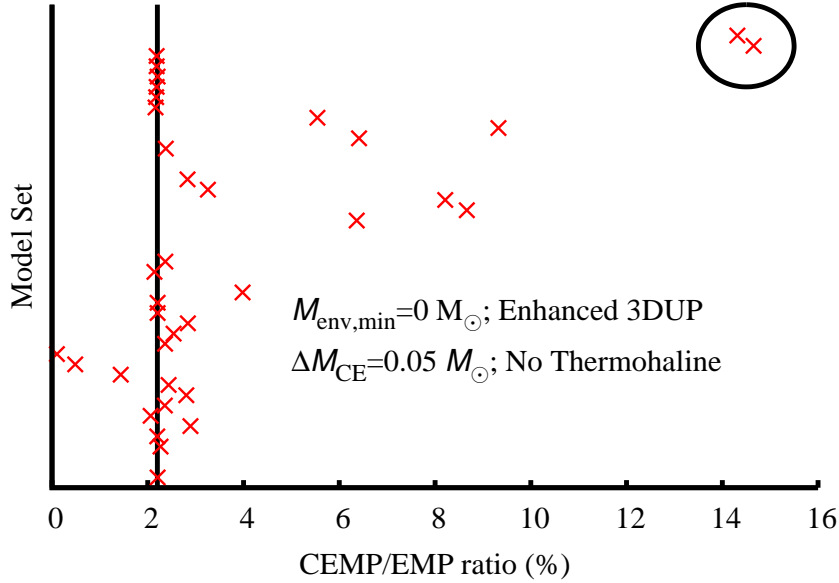}
\end{minipage}
\vspace{8mm} 
\caption[CEMP/EMP ratio from various population synthesis models.]{
  The ratio of carbon-enhanced extremely
metal-poor (CEMP) stars to all extremely metal-poor stars (EMP) in
the Galactic halo as predicted by our population nucleosynthesis code,
\emph{binary\_c}, as a function different input physics, here labelled
'model sets'. Full details of the model sets and general problem
are in \citet{2009A&A...508.1359I}. Briefly, CEMP stars are red giants
that formed by accretion from a carbon-star companion in the past
(\citealp{2015A&A...581A..62A} shows our latest models). Each model
set predicts a CEMP/EMP ratio which should to be compared to the observed
ratio in the Galactic halo which is 20\%. Most of our model sets predict
a $2\%$ CEMP/EMP ratio which is wrong by an order of magnitude. Only
two model sets approach the observed 20\%. These both require all
the following. 1) Third dredge up must be efficient in low-mass stars,
right down to $0.9\mathrm{\,M_{\odot}}$, at low metallicity (here
$Z=10^{-4}$). 2) Third dredge up must be efficient in stars with
small envelope masses (the minimum envelope mass for third dredge
up, $M_{\mathrm{env,}\mathrm{min}},$ is zero here). 3) A small amount
of accretion occurs during any common-envelope phase ($\Delta M_{\mathrm{CE}}=0.05\mathrm{\,M_{\odot}}$).
4) Thermohaline mixing in accreting stars is suppressed. At the time
of writing the original paper \citeyearpar{2009A&A...508.1359I} it
was considered impossible for $0.9\mathrm{\,M_{\odot}}$ stars to
undergo third dredge up and hence become carbon stars. More recent
works have shown that not only is this possible it is quite likely
\citep[e.g.][]{2010MNRAS.403.1413K}. The population synthesis modelling
results encouraged detailed stellar evolutionists to examine stars
of both low mass and low metallicity in far more detail, resulting
in a far better understanding of the problem (e.g.~\citealp{2014MNRAS.438.1741F,2016A&A...592A..29M,2017arXiv170709434M}).}

\label{IzzardFig:2-CEMP-EMP-ratio}
\end{figure}

There are many fields in which population synthesis
has shown itself to be useful. The colours of galaxies are now accepted
to be influenced by binary stars, for example the UV emission of elliptical
galaxies is influenced by subdwarf-O/B stars which can only form in
binaries \citep{2007MNRAS.380.1098H}. Massive blue straggler stars
form when binaries transfer mass or merge and are common enough to
be measurable in the present day mass function of young stellar clusters
\citep{2014ApJ...780..117S}. Similarly, type II supernovae can be
delayed in binary systems, and predictions have recently been made
that analyse the multi-dimensional parameter space of such objects
in great detail \citep{2017A&A...601A..29Z}. The rates of more exotic
channels, such as long and short gamma-ray bursts, can also only be
predicted by population synthesis \citep{2004MNRAS.348.1215I,2017arXiv171005655S}.
Binary stars affect stellar evolution such that the chemical yields
of elements are altered, e.g.~through type Ia supernova and nova
channels. Population synthesis is required to estimate rates and yields
of such events \citep[e.g.][]{2014A&A...563A..83C} and the effect
of mass transfer on chemistry more generally \citep{DeDonder2002,2017arXiv170905237I}.
Mass transfer also makes chemically peculiar stars. The number and
properties of these \textendash{} such as orbital periods and eccentricities
\textendash{} can be used to better understand stellar interior mixing,
mass transfer, tides and interactions with circumstellar and circumbinary
discs (\citealp{2009A&A...498..489J,2010A&A...523A..10I,2013A&A...551A..50D}
and Fig.~\ref{IzzardFig:2-CEMP-EMP-ratio}).
 
We have also recently improved our \emph{binary\_c} models to combine
them with a model of the Galaxy and Magellanic Clouds to predict the
number of hypervelocity stars that originate from the Large Magellanic
Cloud \citep{2017MNRAS.469.2151B}. These young stars are moving so
fast that they are not bound to our Galaxy. One way they can be made
is in binary systems in which one star explodes as a supernova leaving
the companion to escape. We predicted large numbers of such stars,
their positions on the sky and their stellar properties. Our predictions
will soon be tested by \emph{Gaia}.

The recent advances in gravitational wave physics have certainly put
population synthesis in the headlines. Merging stellar-mass black
holes and neutron stars are all the rage, and the predictions of population
synthesis models of such objects can be put to the test \citep[e.g.][and many other works]{2016Natur.534..512B}.
The systematic uncertainties on such rate estimates are large, often
many orders of magnitude. These realistically reflect our ignorance
of the physical processes involved, especially binary problems such
as common envelope evolution, and their sensitivity to the many model
parameters which are often known only approximately.

\section{\emph{Not} Any Colour You Like\ldots{}}

\label{IzzardSec7}\index{gratuitous pink floyd reference}
A persistent claim we have heard over the years
is that, with population synthesis, you can get whatever you like
out of the models. Indeed, with so many parameters to play with this
may seem at first glance to be true. An important counterargument, however,
is that number statistics can be predicted by population synthesis
\emph{with quantified uncertainties. }It is these uncertainties that
are large and perhaps this is where the 'whatever you like' myth
originates. In such cases it is important to identify the parameters
in the model which lead to the large uncertainties and pin these down,
either by observation, experiment or development of a better theory.
By some this is called progress and, with more and better data, it
is simply no longer possible to get any result one would like.
Also, ignorance of the effect of uncertain parameters, be they initial
distributions of stars or the choice of input physics, does not mean
the error on such model predictions is zero. Population synthesis
at least allows us to quantify some of these systematic errors and
remains a very powerful tool to investigate the evolution of both
single and binary stars in the 21st century.




  \bibliographystyle{cambridgeauthordate}
  \label{refs}

 \copyrightline{} 
 \printindex
    
\end{document}